\documentclass[conference]{IEEEtran}

\IEEEoverridecommandlockouts
\usepackage{graphicx}
\usepackage{latexsym}
\usepackage{amsmath}
\usepackage{float}
\usepackage{xspace}
\usepackage[vlined,ruled]{algorithm2e}
\usepackage{amssymb}
\usepackage{subfigure}
\usepackage{url}
\usepackage{lipsum}
\usepackage{paralist}
\usepackage{fixmath}
\usepackage{rotating}
\usepackage{listings}
\usepackage{booktabs}
\usepackage{threeparttable}
\usepackage{seqsplit}
\usepackage{array}
\usepackage{hyperref}

\makeatletter
\newcommand{\thickhline}{%
    \noalign {\ifnum 0=`}\fi \hrule height 1pt
    \futurelet \reserved@a \@xhline
}
\newcolumntype{"}{@{\hskip\tabcolsep\vrule width 1pt\hskip\tabcolsep}}
\makeatother

\usepackage{color}
\usepackage{textcomp}
\definecolor{listinggray}{gray}{0.9}
\definecolor{lbcolor}{rgb}{0.9,0.9,0.9}
\definecolor{lightgray}{rgb}{.9,.9,.9}
\definecolor{darkgray}{rgb}{.4,.4,.4}
\definecolor{purple}{rgb}{0.65, 0.12, 0.82}
\definecolor{forestgreen}{rgb}{0.13, 0.55, 0.13}

\lstdefinelanguage{JavaScript}{
  keywords={typeof, new, true, false, catch, function, return, null, catch, switch, var, if, in, while, do, else, case, break},
  keywordstyle=\color{blue}\bfseries,
  ndkeywords={class, export, boolean, throw, implements, import},
  ndkeywordstyle=\color{black}\bfseries,
  identifierstyle=\color{black},
  sensitive=false,
  comment=[l]{//},
  morecomment=[s]{/*}{*/},
  commentstyle=\color{forestgreen}\ttfamily,
  stringstyle=\color{purple}\ttfamily,
  morestring=[b]',
  morestring=[b]"
}
\newcommand{\javascript}{{JavaScript}\xspace}
\newcommand{\tool}{{JSNose}\xspace}

\newcommand{\head}[1]{\par\smallskip\noindent\textbf{#1.}}

\newcommand{\code}[1]{{\texttt{\small#1}}}

\newboolean{showcomments}
\setboolean{showcomments}{true}
\ifthenelse{\boolean{showcomments}}
{\newcommand{\nb}[2]{
\fbox{\bfseries\sffamily\scriptsize#1}
{\sf\small$\blacktriangleright$\textit{#2}$\blacktriangleleft$}
}
}
{\newcommand{\nb}[2]{}
}

\begin{document}
\title{
Characterizing JavaScript Security Code Smells
}

\author{
\IEEEauthorblockN{Vikas Kambhampati}
\IEEEauthorblockA{New York Institute of Technology\\
Vancouver, BC, Canada\\
vkambh02@nyit.edu}
\and
\IEEEauthorblockN{Nehaz Hussain Mohammed}
\IEEEauthorblockA{New York Institute of Technology\\
Vancouver, BC, Canada\\
nmoham36@nyit.edu}
\and
\IEEEauthorblockN{Amin Milani Fard}
\IEEEauthorblockA{New York Institute of Technology\\
Vancouver, BC, Canada\\
amilanif@nyit.edu}}

%\author{
%\IEEEauthorblockN{Anonymous Authors}
%\IEEEauthorblockA{Anonymous Institute\\
%Anonymous Place\\
%anonymous@anonymous}
%}

\maketitle

\begin{abstract}

\javascript has been consistently among the most popular programming languages in the past decade. However, its dynamic, weakly-typed, and asynchronous nature can make it challenging to write maintainable code for developers without in-depth knowledge of the language. Consequently, many \javascript applications tend to contain code smells that adversely influence program comprehension, maintenance, and debugging. Due to the widespread usage of \javascript, code security is an important matter. While \javascript code smells and detection techniques have been studied in the past, current work on security smell detection for \javascript is scarce. Security code smells are coding patterns indicative of potential vulnerabilities or security weaknesses. Identifying security code smells can help developers to focus on areas where additional security measures may be needed. We present a set of 24 \javascript security code smells, map them to a possible security awareness defined by Common Weakness Enumeration (CWE), explain possible refactoring, and explain our detection mechanism. We implement our security code smell detection on top of an existing open source tool that was proposed to detect general code smells in \javascript. 

\end{abstract}

\begin{IEEEkeywords}
\javascript, security, code smell, smell detection, web applications, refactoring
\end{IEEEkeywords}

\section{Introduction}
\label{Sec:Introduction}

%https://survey.stackoverflow.co/2023/\#technology-most-popular-technologies

According to the Stack Overflow Developer Survey\footnote{\url{https://survey.stackoverflow.co/2024/technology}}, JavaScript has remained the most commonly-used programming language for the 12$^{th}$ year in a row. This is not surprising for a powerful and flexible language that is used for full-stack development of responsive modern web applications. However, its dynamic, weakly-typed, and asynchronous nature, and the complex and dynamic interactions with the Document Object Model (DOM), can make it challenging to analyze, test, and write maintainable code for developers without in-depth knowledge of \javascript \cite{fard2017javascript,amin:ase15}. Hence, applications written in \javascript tend to contain many \emph{code smells}, i.e., patterns in the code that indicate potential comprehension and maintenance issues \cite{fowler1999refactoring}. Detecting JavaScript code smells is important for improving code maintainability, performance, security, and adherence to best practices. %It leads to cleaner and more efficient JavaScript codebases, reduces technical debt, and enhances the overall quality of web applications. 
After detecting code smells, the next steps involve code refactoring such as restructuring or removing the smell. Manual code smell detection is time consuming and error-prone; hence, automated smell detection tools are needed.

\textit{Security code smells} are recurring coding patterns that are indicative of potential vulnerabilities or security weaknesses \cite{rahman2019seven}. These smells can be difficult to spot, as they are often subtle and may not be immediately apparent to developers. Identifying security code smells can help developers to focus on areas where additional security measures may be needed. Examples of such security code smells are usage of hard-coded credentials, weak cryptography, and lack of input validation.
%One common security code smell is the use of hard-coded credentials where a developer includes sensitive information such as usernames, passwords, or API keys directly in the code, rather than storing it in a separate configuration file or environment variable. Other examples are the use of weak hashes to store passwords or the lack of input validation or sanitization. %This occurs when a developer fails to properly validate or sanitize user input, allowing attackers to inject malicious code or data into the system. This can create serious security vulnerabilities, as it allows attackers to execute arbitrary code or access sensitive data. 

While code smells have been extensively studied in the past, not much work has focused on security code smells. Some traditional code smells are known to be related to code security, such as Complex Methods, Long Methods, and Long Parameter Lists \cite{mccabe,alenezi2020relationship,sariman2016novel,singh2020long}. Studies have shown the positive effect of removing code smells on software security \cite{mumtaz2018empirical,ghaith2012improving,Iannone}. A common approach to security smell detection is performing code review that is time-consuming and cumbersome. Automated techniques for detecting these smells can help to fix vulnerabilities before they are established and consequently improve the security of web applications and protect them against potential attacks.

\head{Contributions} \javascript code smells have been studied in the past, however, current work on security smells for \javascript is scarce. In this paper, we propose a list of 24 security code smells for \javascript applications and explain detection methods that have been implemented in the existing JSNose \cite{fard2013jsnose} open source \javascript code smell detector. Milani Fard and Mesbah \cite{fard2013jsnose} explained 13 JavaScript code smells and developed JSNose to detect them \footnote{The work received the Most Influential Paper Award at the 23rd IEEE International Conference on Source Code Analysis and Manipulation in 2023 for its impact on the research community.}.
At the time JSNose was published, server-side \javascript was not very popular and therefore it only focused on client-side \javascript executed in a browser. In this work, we consider 8 of the code smells mentioned in \cite{fard2013jsnose} also as security code smells and present 16 more security smells for \javascript both at client-side and server-side.

Our work makes the following main contributions:
\begin{itemize}

\item We propose a list of 24 \javascript security code smells both at client-side and server-side as our main contribution;
\item We map those security smells to security awareness defined by Common Weakness Enumeration (CWE) \cite{MITRE} and OWASP Top 10 \cite{OWASP};
\item We extend the open source JSNose \cite{fard2013jsnose} tool to detect the proposed \javascript security smells.

\end{itemize}

\head{Outline} This paper is organized as follows. Section \ref{relatedSection} reviews the related work. Section \ref{smellsSection} describes the set of proposed security code smells for \javascript and section \ref{detection} explains detection mechanisms for those smells. Section \ref{discussionSection} explains discussion on our proposed code smells and tool implementation. Finally, Section \ref{conclusionSection} draws conclusions and mentions future works.

\section{Related Work}
\label{relatedSection}

%\subsection{Security Code Smells}
\head{Security Code Smells} While code smells have been extensively studied in the past, not much work has focused on security smells. Some traditional code smells, such as Complex Methods, Long Methods, and Long Parameter Lists are known to be related to code security \cite{mccabe,alenezi2020relationship,sariman2016novel,singh2020long}. Studies have shown the positive effect of removing code smells on security enhancement \cite{mumtaz2018empirical,ghaith2012improving,Iannone}. Paramitha et al. \cite{paramitha2021mining} studied security smell in Java code and found that security smells are correlated with LOC, commit count, and author count. Sultana et al. \cite{sultana2020examining} reported that the smells God Class, Complex Class, Large Class, Feature Envy, Long Parameter List, and Brain Class are correlated with software vulnerabilities, which are also supported by other studies \cite{elkhail2019relating,mumtaz2018empirical,Ghafari,rahman2019seven,rahman2019share}. Ghafari et al. \cite{Ghafari} studied security code smells in Android apps, such as Insufficient Attack Protection, Security Invalidation, Broken Access Control, Sensitive Data Exposure, and Lax Input Validation. Rahman et al. \cite{rahman2019seven, rahman2020code} identified security smells and development anti-patterns that are indicative of security weaknesses in Infrastructure as Code (IaC) scripts including admin privilege by default, empty passwords, hardcoded secrets, invalid IP binding, suspicious comments, using HTTP without TLS, and weak cryptography. In another study, Rahman et al. \cite{rahman2021security} observed that security smells are prevalent in Ansible and Chef as example of IaC scripts. Ponce et al. \cite{ponce2022smells} conducted a literature review on security smells and their impact on microservices applications.

%On relating code smells to security vulnerabilities \cite{elkhail2019relating}

%There was also another research by Rahman et al. which also did literature research to identify security smells in Python Gist [8].

%The article \cite{rahman2019seven} documented the seven different smells in IaC. These smells include 1) granting admin privileges by default, 2) empty passwords, 3) hardcoded secrets, 4) invalid IP address binding, 5) suspicious comments (such as TODO or FIXME), 6) use of HTTP without TLS and 7) use of weak cryptography algorithms. 

%The authors in \cite{Ghafari} performed a literature review on detecting security smells in Android and created a static analyzer to identifies the security smells. 

%\subsection{\javascript Code Smells}
\head{\javascript Code Smells} Commonly used static analysis tools for \javascript, such as ESLint \cite{ESLint}, JSLint \cite{JSLint}, and JSHint \cite{JSHint}, that check the syntax against best practices, can miss security flaws that are potentially security smells. %Moreover, while such linters have been used effectively in real-world applications, their effectiveness is limited by the need to approximate runtime behavior, making precise code smell detection difficult.
%However, detecting code smells in \javascript is challenging due to its dynamic nature. To address this issue, JSNose \cite{fard2013jsnose} and DLint (Gong et al. 2015b) dynamically detect code smells missed by existing static linters. Milani Fard and Mesbah \cite{fard2013jsnose} proposed 13 \javascript code smells (7 generic smells and 6 \javascript-related) and developed to an open source tool, called JSNose, to detect them suing both static and dynamic analysis. JSNose employs a metric-based mechanism to detect code smells at the level of closures, objects, and functions. Examples of such smells are long closure chains, large functions, excessive use of global variables, and potentially dead code. JSNose is flexible to be extended to spot other smells. DLint (Gong et al. 2015b) detects violations of coding practices at the operations level, such as reads and writes on variables and object properties, and function calls. The approach monitors these operations and detects instances of bad coding practices at runtime. Gong et al. experiments suggests that dynamic checking complements static checkers such as JSHint. TypeDevil (Pradel et al. 2015) detects bad practice of type inconsistencies in \javascript  that is missed by generic coding rules checkers such as DLint. SonarQube [5] performs static and dynamic code analysis to detect vulnerabilities, bugs, and code smells. However, more vulnerabilities need to be added to detection of security code smells.
JSNose \cite{fard2013jsnose} and DLint \cite{gong2015dlint} dynamically detect code smells that were not the focus of existing static linters. Milani Fard and Mesbah \cite{fard2013jsnose} proposed 13 \javascript code smells and developed to an open source tool, called JSNose, to detect them using static and dynamic analysis. Examples of such smells are long closure chains, large object, excessive use of global variables, and potentially dead code. JSNose is flexible to be extended to spot other smells. DLint detects violations of coding practices at the operations level, such as reads and writes on variables and object properties, and function calls. TypeDevil \cite{pradel2015typedevil} detects bad practice of type inconsistencies in \javascript that is missed by generic coding rules checkers such as DLint. SonarQube \cite{SonarSource} performs static and dynamic code analysis to detect vulnerabilities, bugs, and code smells. However, more vulnerabilities need to be added to detection of security code smells. Saboury et al. \cite{saboury2017empirical} investigated code smells in JavaScript server-side applications and found that files without code smells have hazard rates lower than files with code smells. Also developers considered some code smells to be serious design problems that hinder the maintainability and reliability of applications.

%In a different context but related to JavaScript, Jafari et al. \cite{jafari2021dependency} examined dependency smells as recurring violations of dependency management guidelines with negative consequences on the project and the ecosystem. These smells are generally introduced as developers react to dependency misbehaviour and the shortcomings of the npm ecosystem.

Previous works have not focused on automatic detection of security smells in \javascript. In this work we study a list of security code smells in \javascript, propose their detection techniques, and implement them within JSNose. To the best of our knowledge there is no widely-used tool that is particularly focused on detecting \javascript security smells.

\section{\javascript Security Code Smells}\label{smellsSection}

Security code smells are coding patterns that could indicate potential future vulnerabilities and weaknesses \cite{Ghafari,rahman2019seven}. Studies in \cite{sultana2020examining,elkhail2019relating,mumtaz2018empirical,Ghafari,rahman2019seven,rahman2019share} have shown that code smells such as God Class, Complex Class, Large Class, Feature Envy, Long Parameter List, and Brain Class are correlated with software vulnerabilities. Complex code is also shown to be related to vulnerabilities; hence, complexity metrics could be used to predict them \cite{shin2010evaluating,shin2008empirical}. The authors in \cite{fard2013jsnose} presented an early work on a collection of \javascript code smells and a tool called JSNose, to detect them. 

\head{Methodology} We began by looking into \javascript code smells mentioned in \cite{fard2013jsnose} that we believe based on the literature can also be considered as security code smells. At the time JSNose \cite{fard2013jsnose} was presented, server-side \javascript was not very popular and therefore it only focused on client-side \javascript executed in a browser. We consider 8 of the code smells in \cite{fard2013jsnose} as security code smells, including {Large Objects}, {Long Method/Function}, {Long Parameter List}, {Empty Catch Blocks}, {Unused/dead code}, {Nested Callback}, {Excessive Global Variables}, and {Coupling between \javascript, HTML, and CSS}.

In addition, we present 16 more security smells for \javascript both at client-side and server-side, making it a total of 24 security code smells. Identification of these smells were based on the literature and our experience. We further map each security smell to a security weakness defined by CWE \cite{MITRE} and OWASP Top 10 vulnerabilities category \cite{OWASP}. We performed these mappings manually and to mitigate error or bias, we conducted a review process in which each author reviewed the mapping done by others. Table \ref{smells} presents our proposed set of security code smells in \javascript and these mappings.

\subsection{Client-side and Server-side Security Smells}

In this section we explain smells that can be found both at the client-side and the server-side \javascript code.

\head{1. Large Objects} God Class, Complex Class, and Large Class, are correlated with vulnerabilities \cite{sultana2020examining,elkhail2019relating,mumtaz2018empirical,Ghafari,rahman2019seven,rahman2019share} and can be considered as security smells. In \javascript as a class-free language, large objects are analogous to large classes. A \javascript object that is doing too much work may be broken into smaller objects to reduce the complexity \cite{fard2013jsnose} and the possibility of hard to detect security issues and weaknesses.

%CWE-1120 (Excessive Code Complexity)
%The code is too complex, as calculated using a well-defined, quantitative measure.
%This issue makes it more difficult to understand and/or maintain the product, which indirectly affects security by making it more difficult or time-consuming to find and/or fix vulnerabilities. It also might make it easier to introduce vulnerabilities. 

%CWE-1093 (Excessively Complex Data Representation)
%The product uses an unnecessarily complex internal representation for its data structures or interrelationships between those structures. This issue makes it more difficult to understand or maintain the product, which indirectly affects security by making it more difficult or time-consuming to find and/or fix vulnerabilities. It also might make it easier to introduce vulnerabilities.

%CWE-506: Embedded Malicious Code

\head{2. Long Method/Function} Long functions are signs of inadequate decomposition and are harder to understand and maintain. More lines of code can introduce more security bugs \cite{mccabe}. Refactoring can be done by splitting into smaller functions. According to CWE-1080 (Source Code File with Excessive Number of Lines of Code), this indirectly affects security by making it more difficult or time-consuming to find and/or fix vulnerabilities. It also might make it easier to introduce vulnerabilities. %While the interpretation of "too many lines of code" may vary for each product or developer, CISQ recommends a default threshold value of 1000.

\head{3. Long Parameter List} A long list of parameters makes the object more complex and harder to maintain and is correlated with vulnerabilities \cite{sultana2020examining,elkhail2019relating,mumtaz2018empirical,Ghafari,rahman2019seven,rahman2019share}. Parameters can be reduced by using objects to combine them \cite{fard2013jsnose}.

\head{4. Empty Catch Blocks} This smell pertains to a poor understanding of the logic in the try block. We can also consider it as a security smell as it opens the possibility of external attack errors/exceptions stay undetected \cite{rahman2019share}.

\head{5. Unused/dead Code} Dead code is code that is never executed. Examples are conditional statements in which the condition will never be satisfied, functions that are never called, unreachable code after an unconditional \code{return} statement \cite{fard2013jsnose}. Such code not only make it difficult to understand but also increase the attack surface and security risk \cite{haas2020static}, and make it more difficult to detect. Due to the dynamic nature of \javascript, an unused code can be suddenly used at runtime. For example an unreferenced function can be called via server-side generated \javascript code or through \code{eval} function use. An adversary may exploit vulnerabilities in such dead code as well. For example, a SQL or command injection in dead code may be exploited to retrieve unauthorized data or to delete data.

\head{6. Nested Callback} Callback functions are passed as arguments to parent functions and are executed after the execution of parent functions. The authors in \cite{fard2013jsnose} considered excessive nested callbacks a \javascript code smell as they make the code hard to read and maintain. A recommended code refactoring is splitting functions and passing references to other functions. We consider such nested callbacks as security smells since according to CWE-1124 (Excessively Deep Nesting), callable code in deep nesting/branching makes it more challenging to maintain and affects security by making it more difficult or time-consuming to find and fix vulnerabilities. It also might make it easier to introduce vulnerabilities \cite{MITRE}.

\head{7. Excessive Global Variables} While the authors in \cite{fard2013jsnose} considered excessive usage of global variables a maintainability and dependability issue, it can also be a security risk. For example, an attacker can extract sensitive data stored in global variables. Changes to global variables can be done from any part of the script even different files loaded on the same page, and hence potentially insecure. Node.js applications often contain this smell. Such excessive usage of global variables violates the \textit{least privilege} principle of security engineering that discourages giving someone more access than they need. All functions can read and write global variables and hence hijacking them would expand the attack surface. JavaScript global identifier conflicts, such as variable type or value conflicts caused by (third-party) script overwrite, may introduce potential risks leading to runtime exceptions and behavior deviation \cite{zhang2020detecting}.

\head{8. Hard-coded Sensitive Information} Writing sensitive information such as usernames, passwords, private keys, or API keys hard-coded in a source code is a security smell, as it makes it easier for attackers to access this information and compromise the system \cite{Ghafari,rahman2019share}. According to CWE \cite{MITRE}, "If hard-coded passwords are used, it is almost certain that malicious users will gain access to the account in question". Hard-coded secrets may not be enough to cause a security breach, thus  this practice is a security smell and not a vulnerability. Instead, such information should be stored in a separate configuration file or environment variable. %This smell is related to CWE-798 (Use of Hard-coded Credentials), CWE-259 (Use of Hard-coded Passwords), and CWE-693 (Protection Mechanism Failure) \cite{MITRE}. Sensitive information which are considered as security smell if written hard-coded include: key, password, secret, and username.
%OWASP Top 10 2021 Category A7 - Identification and Authentication Failures
%OWASP Top 10 2017 Category A2 - Broken Authentication

%Hard-coded passwords in software artifacts is considered as a software security weakness ('CWE-798: Use of Hard-coded Credentials') by Common Weakness Enumerator(CWE)\cite{MITRE}. %According to CWE \cite{MITRE}, "If hard-coded passwords are used, it is almost certain that malicious users will gain access to the account in question".

%Hard-coded credentials are security-sensitive
%https://rules.sonarsource.com/javascript/type/Security%20Hotspot/RSPEC-2068/

\head{9. Dynamic Code Execution} The \code{eval()} function in \javascript allows developers to create and execute new code at runtime. While this can be useful in certain situations, it can also create security vulnerabilities if used improperly. For example, attackers may inject malicious code as user-provided input into the system and execute it through \code{eval()} and perform DOM Cross-Site Scripting (XSS). Using \code{eval} is considered evil and is suggested not be used at all. Evalorizer \cite{meawad2012eval} replaces unnecessary uses of the \code{eval} function with safer alternatives. Evalorizer dynamically intercepts arguments passed to \code{eval} and transforms the \code{eval()} call to a statement or expression without \code{eval}, based on a set of rules. %The approach assumes that a call site of \code{eval()} always receives the same or very similar JavaScript code as its argument.

%***** Dynamic code execution should not be vulnerable to injection attacks
%%%%%%% Ref => https://rules.sonarsource.com/javascript/type/Vulnerability/RSPEC-5334/

%****** Dynamically executing code is security-sensitive
%%%%%%% Ref => https://rules.sonarsource.com/javascript/type/Security%20Hotspot/RSPEC-1523/

%For example, as shown in the following code \code{eval()} evaluates the give string as if it is actual code and produces the output.

%One of the first XSS attack was created by Samy, In 2005, Samy[9] created the first self-propagating XSS worm known as "Samy Worm" or "My Space Worm". Samy worm flooded the My Space, this XSS worm was used by Samy on My Space added over 1 million friends in less than 24 hours. This script force anyone who visited Samy account add Samy as friend including adding text “But Samy is my biggest hero” and eventually copying code to visitors profile[10]. Later it was evident that \code{eval()} was used to evade, circumvent certain limitation to make Samy worm effective to carry the attack [11].

%Using \code{eval()} function compromises integrity of CIA triad if exploited. As a result, \code{eval()} function must not be used at all. With tremendous development in HTML5 and WebAPI most modern browsers can be given instruction to restrict the use of unsafe-eval in manifest. According to OWASP Top Ten[12] using \code{eval()} is the risk of insecure Deserialization. Having this vulnerability there can be threat of different attacks such as code injection, replay attacks, and privilege escalation attack etc.

\head{10. Missing Default in Case Statement} This smell is the recurring pattern of not handling all input combinations when implementing a case conditional logic. Because of this coding pattern, an attacker can guess a value, which is not handled by the case conditional statements and trigger an error. Such an error can provide the attacker unauthorized information for the system in terms of stack traces or system error. 
%\amin{"Such an error can provide the attacker unauthorized information for the system in terms of stack traces or system error.". In the context of client-side JavaScript, how is this possible? Please provide a case in a citation or footnote.}
%
%Let us assume a case where an application processes user input with a switch statement that does not have a default in it. If a user inputs unexpected value, it does not go through any of the developer logic, but bypasses all the cases.
% 
%Example scenario:
% function exampleFn(userInput) {
%     switch (userInput) {
%         case ‘case1’:
%             console.log(‘Case 1 logic’);
%             break;
%         case ‘case2’:
%             console.log(‘Case 2 logic’);
%             break;
%         // No default case
%                 // More logic below that assumes userInput is processed
%     }
% }
% 
% try {
%     exampleFn(userInput);
% } catch (error) {
%     console.error(error); // Could reveal stack traces or other sensitive information
% }
% 
% 
%As seen in the above example, 
%There could be code under the case statements that assumes the input is processed. However, it would not be properly processed, and it could result in unexpected errors and behaviour. 
A malicious actor could persuade the application to behave unpredictably. Without a proper error handling mechanism, it could disclose the stack trace or any other confidential information revealing the application logic. The stack trace could reveal the function name, the location of the error, and the execution flow with line numbers and method hierarchy.
% Example: Error: Data not available
%     at requestingProcess (myProcessor.js:9)
%                                           at requestingProcess (myProcessor.js:4)
%                                           at main.js:98

\subsection{Client-side Security Smells}

\head{11. Coupling between \javascript and HTML} %A group of code smells studied in \cite{fard2013jsnose} are related to the coupling between \javascript, HTML, and CSS. 
Mixing \javascript code with markup can make it difficult to understand, maintain and debug the web application. Similar to \cite{fard2013jsnose} we present security smells in 2 categories:  %\cite{nguyen2012detection,Volkan,JSCamp2009,maintainbleJS}. %Separation is a benefit of unobtrusive DOM scripting described in the \javascript manifesto \cite{JSManifesto}: \textit{all \javascript code should be maintained separately, without impacting other files of script, markup (XHTML) or code (PHP, JSP, or other languages)}. A web application that its JS, CSS, and HTML components are loosely coupled is easier to debug and maintain.

\textit{a) \javascript in HTML.} Inline assignment of event handlers in the HTML code, e.g. \code{<button onclick="foo();" id="myBtn"/>} is a code smell as it tightly couples the HTML code to the \javascript code \cite{fard2013jsnose}. Using inline \javascript makes web applications more vulnerable to Cross-Site Scripting (XSS) attacks. This security risk can be avoided by having all scripts including inline event handlers (e.g. onclick) in external .js files. For better security it is also recommended to establish a Content Security Policy (CSP). This is a security layer in the communication between client and server that allows adding content security rules to HTTP response headers. Moreover, one can use the \code{script-src} and \code{default-src} directives in CSP to block all inline scripts so that no malicious inline script can be executed.

\textit{b) HTML in \javascript.} The \javascript code can manipulate the DOM through DOM APIs and embedded HTML strings in \javascript that is run by the browser. Examples are \code{createElement()}, \code{createTextNode()}, and \code{appendChild()}. While extensive usage of DOM API calls and embedded HTML in \javascript is considered as code smell that can be refactored by moving the HTML code to a template \cite{fard2013jsnose}, it can also be a security risk. An adversary could exploit a vulnerability in frameworks or browsers to execute a malicious code. DOM manipulations require double check to ensure all user-input are sanitized and there are no forbidden tags and malicious code. DOM elements that can attach inline \javascript event handlers should also be checked.

\head{12. Unverified Cross-Origin Communications} Cross-origin communication between Window objects; e.g., between a page and a pop-up, or between a page and an iframe within it, can be done using the \code{window.postMessage()} method \cite{postMessage,SonarSource}. Therefore it is important to verify the identity of sender and receiver. When sending a message with \code{postMessage()} to other windows, the receiver should be defined and using wildcard (*) is not recommended as a malicious site can change the location of the window and intercept the data. Also if you expect to receive messages from other sites, always verify the sender using the origin and possibly source properties. Otherwise any window, such as one browsing a malicious website, can send a message to any other window, and you cannot stop an unknown sender to send malicious messages to any other window. %If you do not expect to receive messages from other sites, do not add any event listeners for message events. 

\head{13. Active Debugging Code} Having debug features remaining active/enabled in the production code is a security concern as they can reveal detailed information of the system/user that runs the application. If the debugging code is not disabled when executing in a production environment, then sensitive information may be exposed to attackers. Debugger statements must be removed from the production code to prevent any vulnerability to attacks. Examples of debugging code in \javascript are the use of \code{console.log()}, \code{console.debug()}, \code{console.error()}, or even \code{alert()}.

%alert(...) as well as confirm(...) and prompt(...) can be useful for debugging during development, but in production mode this kind of pop-up could expose sensitive information to attackers, and should never be displayed.

%A common development practice is to add "back door" code specifically designed for debugging or testing purposes that is not intended to be shipped or deployed with the product. These back door entry points create security risks because they are not considered during design or testing and fall outside of the expected operating conditions of the product \cite{MITRE}.

%An application's debug features enable developers to find bugs more easily and thus facilitate also the work of attackers. 

%"alert(...)" should not be used
%%%%%%% Ref => https://rules.sonarsource.com/javascript/type/Vulnerability/RSPEC-1442/

%******Using hardcoded IP addresses is security-sensitive
%%%%%%% Ref => https://rules.sonarsource.com/javascript/type/Security%20Hotspot/RSPEC-1313/

\begin{table*}[t]
\caption{Proposed security code smells for \javascript. The first 8 are among the \javascript code smells mentioned in \cite{fard2013jsnose}.} \label{smells}
\vspace{-10pt}
{\scriptsize
%{\footnotesize
\begin{center}
\begin{tabular}{p{2.9cm}|p{10.6cm}|p{3.3cm}}%{l|l|l|l}
\hline
 \textbf{Security Code smell} & \textbf{Common Weakness Enumerator} \cite{MITRE} & \textbf{OWASP Top 10} \cite{OWASP} \\
\hline
\hline

Large Object & CWE-1120 (Excessive Code Complexity), CWE-1093 (Excessively Complex Data Representation), CWE-1080 (Source Code File with Excessive Number of Lines of Code) & Insecure Direct Object References \\
 \hline
Long Method/Function & CWE-1080 (Source Code File with Excessive Number of Lines of Code), CWE-1120 (Excessive Code Complexity) & Insecure Direct Object References \\
 \hline
Long Parameter List & CWE-1120 (Excessive Code Complexity), CWE-1093 (Excessively Complex Data Representation) & Injection \\
\hline
 Empty Catch Blocks & CWE-703 (Improper Check or Handling of Exceptional Conditions), CWE-1069 (Empty Exception Block), CWE-1071: Empty Code Block & Improper Error Handling \\
\hline
 Unused/dead code & CWE-561 (Dead Code), CWE-1164 (Irrelevant Code) & Injection \\
\hline
 Nested Callback & CWE-1124 (Excessively Deep Nesting) & Security Misconfiguration \\
 \hline
 Excessive Global Variables & CWE-1108: Excessive Reliance on Global Variables & Insecure Direct Object References \\
 \hline
 Coupling between JS and HTML & CWE-116: Improper Encoding or Escaping of Output, CWE-829: Inclusion of Functionality from Untrusted Control Sphere & Cross-Site Scripting \\
 \hline

 Hard-coded Sensitive Information & CWE-798 (Use of Hard-coded Credentials), CWE-259 (Use of Hard-coded Passwords), and CWE-693 (Protection Mechanism Failure) & Identification and Authentication Failures \\
\hline
 Missing Default in Case Statement &  CWE-478 (Missing Default Case in Switch Statement) & Insecure Direct Object References, Injection  \\
\hline
 Use of Weak Cryptography &  CWE-326 (Inadequate Encryption Strength), CWE-327 (Use of a Broken or Risky Cryptographic Algorithm), CWE-328 (Use of Weak Hash), CWE-1240 (Use of a Risky Cryptographic Primitive)  & Cryptographic Failures \\
\hline
 Insecure HTTP & CWE-319 (Cleartext Transmission of Sensitive Information) & Cryptographic Failures  \\
 \hline
 Unverified Cross-Origin Communications & CWE-345 (Insufficient Verification of Data Authenticity) & Broken Access Control \\
\hline
 Active Debugging Code & CWE-489 (Active Debug Code), CWE-215 (Insertion of Sensitive Information Into Debugging Code) & Sensitive Data Exposure \\
\hline
 Dynamic Code Execution & CWE-95 (Improper Neutralization of Directives in Dynamically Evaluated Code), CWE-77 (Command Injection), CWE-20 (Improper Input Validation) & Injection \\
\hline
 Insecure DOM Manipulation & CWE-79 (Improper Neutralization of Input During Web Page Generation ('Cross-site Scripting')) & Injection \\
\hline
 Unvalidated Redirect & CWE-20 (Improper Input Validation), CWE-601 (URL Redirection to Untrusted Site ('Open Redirect')) & Broken Access Control \\
 \hline
 JSON Injection & CWE-74 (Improper Neutralization of Special Elements in Output Used by a Downstream Component ('Injection')), CWE-116: Improper Encoding or Escaping of Output, CWE-77 (Command Injection) & Injection \\
\hline
 Unprotected Cookies & CWE-614 (Sensitive Cookie in HTTPS Session Without 'Secure' Attribute), CWE-315 (Cleartext Storage of Sensitive Information in a Cookie), CWE-311 (Missing Encryption of Sensitive Data), CWE-565 (Reliance on Cookies without Validation and Integrity Checking) & Insecure Design, Security Misconfiguration   \\
\hline
 Long Prototype Chain & CWE-1074 (Class with Excessively Deep Inheritance) & Injection \\
 \hline
 Prototype Pollution & CWE-1321 (Improperly Controlled Modification of Object Prototype Attributes ('Prototype Pollution')) & Cross-Site Scripting \\
\hline

% Insecure Dependencies & CWE-1395 (Dependency on Vulnerable Third-Party Component), CWE-1104 (Use of Unmaintained Third Party Components) & vulnerable and outdated components \\
% \hline

Logging Sensitive Information & CWE-532 (Insertion of Sensitive Information into Log File), CWE-200 (Exposure of Sensitive Information to an Unauthorized Actor), CWE-312 (Cleartext Storage of Sensitive Information) & Security Logging and Monitoring Failures \\
\hline

Insecure File Handling & CWE-434 (Unrestricted Upload of File with Dangerous Type), CWE-646 (Reliance on File Name or Extension of Externally-Supplied File) & Insecure Data Storage \\
\hline

Error Handling Disclosure & CWE-209 (Generation of Error Message Containing Sensitive Information), CWE-497 (Exposure of Sensitive System Information to an Unauthorized Control Sphere) & Improper Error Handling \\
\hline

% \hline
\end{tabular}
\end{center}
}
%\vspace{-6pt}
\end{table*}

\head{14. Insecure DOM Manipulation} 
%\textit{a) Using \code{Element.innerHTML}} 
In JavaScript the \code{innerHTML} property allows developers to set the HTML content of an element. While this helps to manipulate the DOM, it can also create vulnerabilities if used improperly. For example, if a developer uses \code{innerHTML} to set the content of an element based on user-provided input, it could allow attackers to inject malicious code into the page. Using \code{innerHTML} or \code{outerHTML} even for setting text content can be dangerous. Using \code{innerHTML} with unsafe or unescaped text can lead to the DOM XSS. Thus, whenever it is possible it is better to use \code{innerText} or \code{textContent}. %%For example, profile picture is being set on web application using \code{innerHTML} on div, following malicious code masquerades as image tag.
% \begin{lstlisting}
% <img onerror='window.eleid.remove()' src='invalid-image'>
% \end{lstlisting}
% As it can be noticed that, invalid image source is specified in the above image tag. When given image is failed to load, \code{onerror} callback function will be executed. Thus, whenever this image is loaded malicious code specified in the \code{onerror} handler. It is also possible for attacker to perform CSRF (Cross site request forger) attack.
%As stated in the OWASP Top Ten [12] cross site scripting (XSS) is one of the top 10 security risks of web applications. If this vulnerability is unattended, threat agent can exploit it to further compromise the system even can perform other attacks such as Cross site request forgery (CSRF).
DOM XSS happens when user-controlled data is used in the \javascript code. User-controlled data should always be validated and should consider as untrusted. Consider the URL \code{\seqsplit{http://app/page.html\#<img onerror="document.write('hacked')";src='invalid-image'/>}} that is used to set the \code{innerHTML} of a DOM element. An attacker may pass malicious code in fragment to perform DOM XSS. In this scenario, malicious code that manipulate the DOM is masqueraded as image.
%Taking untrusted user-controlled data and setting \code{innerHTML} can result in DOM-XSS:
% \begin{lstlisting}
% const myElement = document.getElementById('myElement');
% const hash = decodeURIComponent(location.hash.substr(1));
% myElement.innerHTML = hash;
% \end{lstlisting}
Possible refactoring of this smell is either not to use \code{innerHTML} or escape unsafe character such as $<$, $>$, and \& before setting \code{innerHTML}. %Above code is refactored by replacing \code{innerHTML} with \code{innerText} as follows

%https://stackoverflow.com/questions/63159087/how-to-work-around-setting-innerhtml-causing-escape-sequences-to-expand

% \begin{lstlisting}
% const myElement = document.getElementById('myElement'); 
% const hash = decodeURIComponent(location.hash.substr(1)); 
% myElement.innerText = hash;
% \end{lstlisting}

%\textit{b) Use of \code{document.write()}} 
Another potential insecure DOM manipulation is the use of \code{document.write()} that allows developers to write content to the current document. While this helps with generating dynamic content, it can also create security vulnerabilities. For example, if a developer uses \code{document.write()} to write user-provided input to the page, it could allow attackers to inject malicious content into the document. Such input should always be untrusted and sanitized first.

\head{15. Unvalidated Redirect} %Unvalidated redirects occur when a web application takes the user-controlled untrusted input data that could potentially redirect requests to the URL specified in the insecure input [13]. 
Unvalidated redirect occurs when a web application takes untrusted input data from an attacker that results in redirecting a user to a malicious website in attempt to steal the user data for example credentials or personal identifiable information. For example an adversary can craft the URL to redirect the victim to the malicious website such as:

http://www.example.com/page?url=malicious.example.com

\head{16. JSON Injection} JSON injection happens when untrusted malicious data is injected into a JSON string without sufficient validation or sanitization. 
Here is an illustration of a client-side JSON injection attack. %The starting JSON string is identical to the one from the preceding example.
The server does not sanitize before retrieving the JSON data from an unreliable source, which may include a malicious payload. The JSON string is parsed by the client using the \code{eval()} function:

\begin{lstlisting} 
var result = eval("(" + json_string + ")");
document.getElementById("#accountType").innerText = result.account;
document.getElementById("#userName").innerText = result.name;
document.getElementById("#pass").innerText = result.pass;
\end{lstlisting} 

The attacker injects the following value for \code{accountType}:

\begin{lstlisting}
user"});alert(document.cookie);({"accountType":"user
\end{lstlisting} 

The eval function executes the alert call.
Parsing the manipulated string leads to a XSS attack, resulting in the disclosure of \code{document.cookie}. 
%{What are the measures to mitigate JSON injection attacks?}
%Early integration of security controls is essential to prevent the potential damage that server-side JSON injections may create. 
To mitigate JSON injection we should prevent the loading of sensitive data, such as account information and rights, from JSON data controlled by users or from unreliable sources. Content Security Policy (CSP) can be to imposed to prevent client-side JSON injections by forbidding the usage of \code{eval()} and forcing developers to choose the more secure \code{JSON.parse} function instead. %This makes the application's processing of JSON data more secure.

\head{17. Unprotected Cookies} HTTP cookies are frequently used by web browsers for user authentication and session persistence. Cookies can be used maliciously by attackers. When a cookie is secured with the secure attribute set to \code{true}, the browser would not send it over an unencrypted HTTP request. %, making it impossible for a man-in-the-middle attacker to view it. %Cookies and the malicious actions mentioned above are both common components of these three attacks: Cross-site scripting (XSS), man-in-the-middle attacks (MITM), and Cross-site request forgery attacks (CSRF).
% \begin{lstlisting}
% Cookie c = new Cookie(COOKIENAME, sensitivedata);
% c.setSecure(false); // This code creates a cookie called "c" with sensitive data and disables the secure flag, since the flag is set to False.
% \end{lstlisting}
%\textbf{Corrective mechanism:} 
Since HTTPS is advised everywhere, setting the secure flag to true when creating cookies ought to be the default action. Also the secure flag should be set to true for session-cookies.

% \begin{lstlisting}
% Cookie c = new Cookie(COOKIENAME, sensitivedata);
% c.setSecure(true); // This code creates a cookie named "c" with sensitive data and ensures that it will only be sent during encrypted HTTPS requests by setting the secure flag to true.
% \end{lstlisting}

%****************** SONR ******************
%****Creating cookies without the "secure" flag is security-sensitive
%%%%%%% Ref => https://rules.sonarsource.com/javascript/type/Security%20Hotspot/RSPEC-2092/
%******************************************

%****** jQuery has a (recently patched) vulnerability to "prototype pollution

%%%%%%% Ref => https://www.oscommerce.com/forums/topic/493360-jquery-vulnerability-upgrade-needed/
%%%%%%% Ref => https://www.darkreading.com/application-security/new-fix-for-jquery-vulnerabilities/a/d-id/750998

%Prototypes are the mechanism by which objects can inherit features from one another. This means that objects can act as a template for other inheriting objects, that will themselves maybe be inherited and so on. This course of action is called a prototype chain. It is the reason why some objects will have properties of other objects and why it is so easy to transfer those properties.

\head{18. Long Prototype Inheritance Chain} While the modern \javascript supports class-based inheritance and concepts such as Mixins, it is originally a class-free prototypal inheritance language, i.e., objects can inherit properties from other objects. Every object in \javascript has a prototype. When a messages reaches an object, \javascript will attempt to find a property in that object first, if it cannot find it then the message will be sent to the object's prototype and so on. The \code{extends} keyword in class-based inheritance also works internally using this old prototype mechanics. Long prototype inheritance chains, specially in scattered files and parts of code, make the code difficult to understand and maintain and also affects the performance in accessing the object property. This indirectly affects security by making it more difficult or time-consuming to find and/or fix vulnerabilities. It also might make it easier to introduce vulnerabilities, as explained in CWE-1074 \cite{MITRE}.

% \begin{lstlisting}
% var a = {p1: 1, p2: "test"};
% var b = Object.create(a);
% var c = Object.create(b);
% var d = Object.create(c);
% // d --> c --> b --> a --> Object.prototype --> null
% \end{lstlisting}

%From: https://learn.snyk.io/lessons/prototype-pollution/javascript/

\head{19. Prototype Pollution} Prototype pollution is a vulnerability that enables malicious actors to inject values that overwrite or corrupt the "prototype" of a base object. The malicious prototype can propagate to numerous other objects that inherit it. By gaining control over the default values of an object's properties, attackers can manipulate the application's logic resulting in Denial-of-Service or Remote Code Execution.

On web browsers, prototype pollution commonly leads to XSS attacks. %In 2019, for instance, a prototype pollution bug found in the JavaScript library jQuery left many web applications vulnerable to such assaults.
%As prototype pollution lets you control properties that would otherwise be inaccessible, this potentially enables you to reach a number of additional sinks within the target application. 
Developers who are unfamiliar with prototype pollution may wrongly assume that these properties are not user controllable, which means there may only be minimal filtering or sanitization in place. 
For instance, when user-controllable objects are obtained from a JSON string using \code{JSON.parse()} method, it is important to note that \code{JSON.parse()} treats all keys in the JSON object as arbitrary strings. This includes special keys like "{\_}{\_}proto{\_}{\_}". This behaviour opens the possibility of prototype pollution, where an attacker can inject malicious JSON data, such as through a web message, and potentially compromise the application's prototype chain.
% \begin{lstlisting}
% {
%     "__proto__": {
%         "maliciousProperty": "payload"
%     }
% }
% \end{lstlisting}
If you use the \code{JSON.parse()} function to turn this into a JavaScript object, the resultant object will have a property with the key {\_}{\_}proto{\_}{\_}:

\begin{lstlisting}
const objectLiteral = {__proto__: {maliciousProp: 'payload'}};
const objectFromJson = JSON.parse('{"__proto__": {"maliciousProp": "payload"}}');
objectLiteral.hasOwnProperty('__proto__');     // false
objectFromJson.hasOwnProperty('__proto__');    // true
\end{lstlisting}

The assignment will also result in prototype pollution if the object produced by \code{JSON.parse()} is later merged into an already-existing object without adequate key sanitization.
%\textbf{Corrective mechanisms:} 
When you need to recursively set nested properties on an object, use well-known open-source libraries to reduce prototype pollution vulnerabilities in your codebase. Always check for vulnerabilities in the library you choose. Use \code{Object.create(null)} to completely avoid using prototypes or \code{Object.freeze(Object.prototype)} to stop any alterations to the shared prototype to secure your code further.

\subsection{Server-side Security Smells}

\head{20. Use of Weak Cryptography} Using weak cryptography algorithm is categorized as security smell \cite{rahman2019share} %This smell is the recurring pattern of using weak encryption and weak hash algorithms. %Using weak cryptography algorithms for setting up passwords may lead to a breach. 
%Many cryptographic algorithms provided by cryptography libraries are known to be weak, or flawed. %Using such an algorithm means that encrypted or hashed data is less secure than it appears to be. 
that may result in sensitive data exposure, key leakage, broken authentication, insecure session, and spoofing attacks \cite{SonarSource}. Some weak encryption algorithms AES-ECB, DES, RC2, and RC4, are not recommended to be used. Instead, modern strong cryptographic algorithm such as AES-128 or RSA-2048 should be used for encryption. Cryptographic hash algorithms such as MD2, MD4, MD5, MD6, HAVAL-128, HMAC-MD5, DSA, RIPEMD, RIPEMD-128, RIPEMD-160, and SHA-1 are not secure, because are susceptible to collision attacks and modular differential attacks and it is not difficult to find two or more different inputs that produce the same hash. Safer alternatives, such as SHA-256, SHA-512, and SHA-3 are recommended.

\head{21. Insecure HTTP} This smell is the recurring pattern of using HTTP without the Transport Layer Security (TLS v1.2 or 1.3) which makes the communication between two entities less secure and susceptible to man-in-the-middle attacks. %For example, as shown in Figure 5, the authentication URL uses HTTP without SSL/TLS for ‘auth url’. Such usage of HTTP can be problematic, as an attacker can eavesdrop on the communication channel. 

\head{22. Logging Sensitive Information} Server-side logging, e.g. by using \code{console.log()} or \code{console.error()}, helps developers fix bugs and enhance code reliability. However, If sensitive data, such as passwords, API keys, personally identifiable information, financial data, full path names, or system information, are written to a log entry it could be exposed to an attacker who gains access to the logs. If such sensitive data needs to be logged for debugging purposes, one can mask or encrypt them.

\head{23. Insecure File Handling} This type of vulnerability occurs when the server does not validate, sanitize, or secure the files that are being uploaded by the users. This makes the server vulnerable to various security vulnerabilities such as the attacker can upload malicious files into the server. Once the attacker is successful in uploading the malicious file they can leverage it to execute arbitrary code on the server.

\head{24. Error Handling Disclosure} This code smell relates to the vulnerability which is caused when an application inadvertently discloses the sensitive information in the error messages. The information disclosed in the error messages can then be leveraged by the attackers such as to gain information on the internal working of the application. Error messages that include stack traces may reveal method names, line numbers, and file paths disclosing the internal file structure and server configuration \cite{gadient2021security}.

\section{Security Smell Detection}\label{detection}

%This section presents mechanisms to detect the proposed security code smells. 

\head{Implementation} While some of the above-mentioned security smells, such as long parameter list, long method/function, or empty catch blocks, can be detected using existing tools such as ESLint, many of them are not targeted by common static code analyzers. In this work, we extend JSNose \cite{fard2013jsnose} to identify our set of security smells. In order to use JSNose to identify our proposed set of security smells, we extended \code{SmellDetector}, \code{JSModifyProxyPlugin}, and \code{Crawler} classes of JSNose \cite{fard2013jsnose} and made the forked repository available for further extensions\footnote{\url{https://github.com/nyit-vancouver/JavaScriptSecurityCodeSmells}}.

The authors of JSNose received the \textit{Most Influential Paper Award} at the \textit{23rd IEEE International Conference on Source Code Analysis and Manipulation in 2023} for its impact on the research community, and that is why we decided to extend their implementation instead of developing our own tool.

\begin{figure}[t]
\centering
\includegraphics[trim = 0mm 3mm 0mm 1mm,width=0.9\hsize]{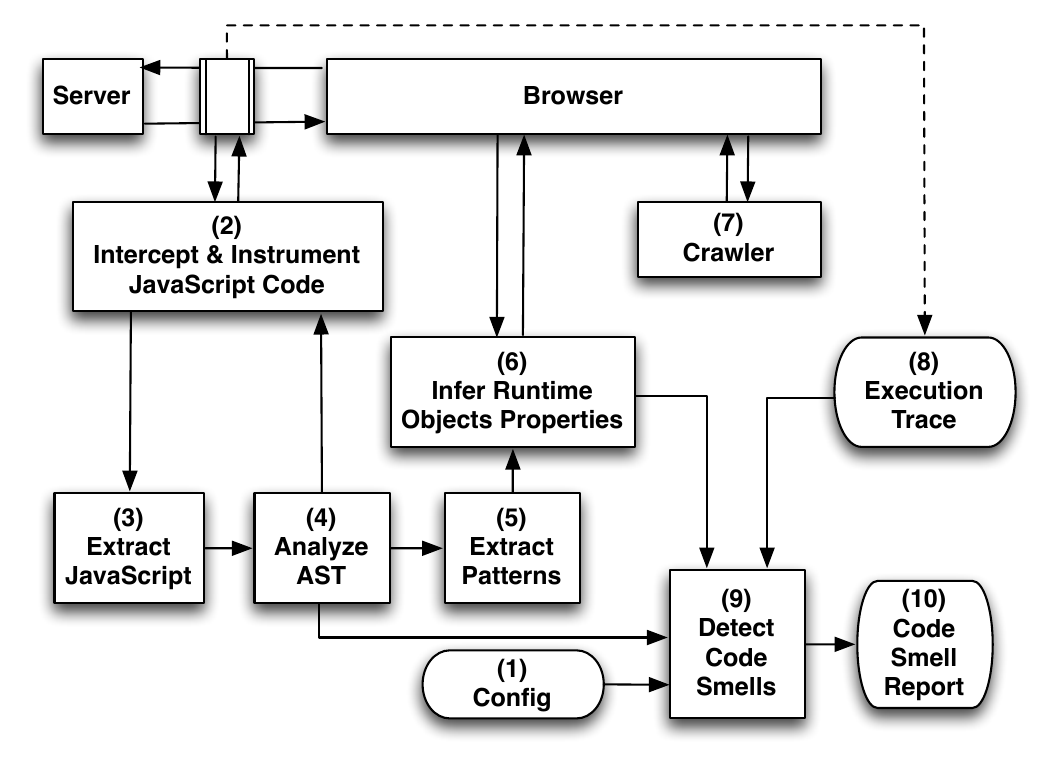}
\caption{Processing view of \tool \cite{fard2013jsnose} for detecting \javascript code smells at client-side.}
 \label{Fig:overview}
\end{figure} 

It is worth mentioning that when JSNose was published, server-side \javascript was not very popular, and therefore, it only focused on client-side \javascript executed in a browser. It applies a metric-based approach to identify smelly sections of the code, i.e., a traditional heuristic-based method of extracting source code metrics and checking against set thresholds \cite{lanza2006objectoriented,moha2010decor,munro2005product,simon2001metrics}. To do so, JSNose extracts objects, functions, and their relationships from the source code using static and dynamic code analysis. Figure \ref{Fig:overview} shows an overview of JSNose, which (1) gets metrics and thresholds as configuration, (2) intercepts the \javascript code of a web application via a proxy between the server and the browser and instruments the code to monitor its execution, (3) extracts the code, (4) parses into an Abstract Syntax Tree (AST), (5) extracts objects, properties, functions, and code blocks, (6) infers dynamic objects, types, and properties from the browser at runtime, (7) uses a crawler to interact with the instrumented application in the browser, (8) collects execution traces, (9) detects code smells based on the defined metrics, and (10) report it in a text file. 

%A sample selection of a report is shown below.

%\begin{tiny}
% \begin{lstlisting}[frame=single,basicstyle=\tiny\ttfamily,language=]
% ***************************************
% ********** CODE SMELL REPORT **********
% ***************************************
% ********** LARGE OBJECT **********
% Number of occurance: 6
% Item: Effect in JS file: showTodos_actionscript12 at line number: 7
% ...
% ********** LONG METHOD/FUNCTION **********
% Number of occurance: 12
% Item:  in JS file: calendar_js at line number: 963
% ...
% ********** LONG PARAMETER LIST **********
% Number of occurance: 2
% Item: commonAddTodo in JS file: showTodos_actionscript11 at line number: 169
% ...
% ********** COUPLING JS/HTML **********
% Total number of JavaScript in HTML tags: 1
% INPUT onclick="document.forms[0].elements['method'].value='update';"
% Occurance of CSS in JavaScript
% Number of occurance: 16
% Item: CSS in JavaScript in JS file: showTodos_actionscript11 at line number: 210
% ...
% ********** EXCESSIVE GLOBAL VARIABLES **********
% Number of global variables: 25
% List of  global variables: [Autocompleter, Field, dwr, 
% ....
% ]
% \end{lstlisting}
% %\end{tiny}

\head{Code Smells Detected by JSNose} For Large Objects, Long Parameter List, Empty Catch Blocks, Unused/Dead Code, Nested Callback, Excessive Global Variables, and Coupling between JS and HTML, we follow the same as \cite{fard2013jsnose} with the default settings. For the Long Method/Function smell, while the interpretation of too many lines of code for CWE-1080 (Source Code File with Excessive Number of Lines of Code) may vary for each product or developer, the Consortium for Information \& Software Quality (CISQ) recommends a threshold of 1000 LOC \cite{CISQ}.

\head{Hard-coded Sensitive Information} We check for some widely used patterns such as authenticated URLs, e.g., a Mongo connection string with 
\code{\seqsplit{"mongodb://[username:password@]host1[:port1],...hostN[:portN]][/[defaultauthdb][?options]]"}} in order to report \code{\seqsplit{"mongodb://username:password@localhost:27017/mydb?authSource=admin\&ssl=true"}}. We also look for common variable names that hold sensitive information such as user, username, uname, password, passwd, pwd, or key. We report such variables as potential security smells if fixed string values are assigned to them.

\head{Missing Default in Case Statement} Once a \code{switch} statement ASTNode is found, we check if the \code{default} keyword is present before a new node data.%, and if not report it as security smell.

\head{Use of Weak Cryptography} We look for the usage of weak encryption algorithms AES-ECB, DES, RC2, and RC4, and also search for the use of weak hash algorithms such as MD2, MD4, MD5, MD6, HAVAL-128, HMAC-MD5, DSA, RIPEMD, RIPEMD-128, RIPEMD-160, and SHA-1.

\head{Insecure HTTP} We get the HTTP protocol that is used by the current page by 
\\\code{window.location.protocol} and check if the communication between the server and client is encrypted.

\head{Unverified Cross-Origin Communication} As the AST of extracted \javascript code is parsed to navigate through the pages, the static analyzer checks for the event listeners and whether the source has been specified or not. This helps in finding out if the communication is established only with the expected external servers. We report all other instances where the external servers are communicated as code smells.

\head{Active Debugging Code} During the static analysis we find the locations where the \code{console.log()}, \code{console.debug()}, \code{console.error()}, and \code{alert()} are used to report them as potential security smells. %The Window object is used by the SmellDetector class for this purpose.

\head{Dynamic Code Execution} %Using \code{eval} is highly discouraged, there is no excuse to use the \code{eval()} function in the JavaScript code. If a threat agent exploits \code{eval()} code injection and privilege escalation are possible. 
During static analysis we check whether \code{eval()} function is present. Precise detection of malicious code as input to \code{eval()} requires dynamic analysis.

\head{Insecure DOM Manipulation} Using \code{innerHTML} with untrusted data can lead to the DOM-XSS. Detection is possible using a static analyzer by checking whether this attribute is being used to send content on an element. If we find an instance of \code{document.write()} during Window object data traversal, we report it as a security smell too.

\head{Unvalidated Redirects} During the AST traversal if redirect is present, we check if redirect is safeguarded by conditional statement. For example, a URL that starts with / or // is safe as it redirects to the same domain. URLs that start with http:// or https:// or ftp:// should be restricted and considered potentially unsafe unless it is redirected to a whitelist domain. \javascript URIs such as \code{javascript:alert()} and Data URIs %such as \code{data:text/html,<script>document.write('hacked')</script>} 
are also considered unsafe by default.

\head{JSON Injection} JSON Injections are due to non-validated data turned into a JSON format and passed to the server. We look for instances of manually forming a JSON structure, e.g., \code{var name = '{"user": "' + inputData + '"}';}. Such examples are considered as an attempt to construct a JSON structure manually instead of using safe methodologies like using built-in JSON serialization functions.

\head{Unprotected Cookies} We report cookies as possible code smells if the retrieved data from the cookies are used without validation. The cookieProtect method of SmellDetector checks for any instances that use the Cookie getter method to determine if any validation mechanism is implemented on the retrieved data. 

\head{Long Prototype Chain} In prototypal inheritance, the object's prototype is linked to another object's prototype. Once the length of the prototype chain exceeds 7, we consider it as a security code smell. This maximum inheritance level of 7 is recommended by the Consortium for Information \& Software Quality (CISQ) Automated Quality Characteristic Measures \cite{CISQ} with regards to CWE-1074.% (Class with Excessively Deep Inheritance).

\head{Prototype Pollution} To detect this smell, initially we look for instances of Object prototype modification as part of static analysis. The input of that modification is further analyzed by reverse engineering using the Window data to see if it underwent any form of validations or sanitizations. If no such occurrences are found, it is reported as a security smell.

\head{Error Handling Disclosure} We look for stack traces in the returned HTTP body with terms related to application crashes such as "stack," "trace," and "error" \cite{gadient2021security}.

\section{Discussion}
\label{discussionSection}

The studied security code smells may be checked during development, code reviews, testing, or at CD/CI workflows to improve code security. While our proposed security smells are mainly for vanilla JavaScript that may not be written much at large-scale these days, they can be adapted to TypeScript as well as popular frameworks such as Angular, VUE, and React. %The paper does not mention them, even though vanilla JavaScript code is rarely written these days (if ever); most of the time, framework-generated code runs on the client side, and most programmers use TypeScript today.

\head{Tool Support} Existing commercial static code analyzers support some of our proposed security code smells. Veracode \cite{Veracode} primarily focuses on identifying security vulnerabilities rather than code smells. It can detect issues such as hard-coded sensitive information, dynamic code execution, and insecure dependencies. However, it does not mention support for detecting all our proposed code smells, such as large objects, empty catch blocks, or nested callback. Snyk \cite{Snyk} detects vulnerabilities in dependencies and containers, insecure dependencies, and prototype pollution. It may not cover all the code smells like large objects, empty catch blocks, or unused code. SonarQube \cite{SonarSource} supports a wide range of code smells and vulnerabilities. It can detect issues such as long method/function, long parameter list, empty catch blocks, unused code, and hard-coded sensitive information. SonarQube is more likely to cover most of the security code smells listed in our paper, making it a suitable tool for comprehensive code quality checks. Checkmarx \cite{Checkmarx} can detect issues related to dynamic code execution, insecure dependencies, hard-coded sensitive information, and other standard security weaknesses. It can identify many code smells mentioned in our paper, though explicit support may vary.

% Summary
% Veracode: Good for security vulnerabilities but not explicitly comprehensive for all proposed code smells.
% Snyk: Excellent for dependency vulnerabilities but limited on general code smells.
% SonarQube: Broad coverage of code smells and security vulnerabilities will likely cover most of your proposed smells.
% Checkmarx: Strong in security vulnerability detection, with a good range of supported code smells.

\head{Implementation and Evaluation} We decided to extend the detector of open-source JSNose \cite{fard2013jsnose} as manipulating the code for static and dynamic analysis is straightforward. We executed the tool on a given simple single-page web application that we implemented ourselves, which contains patterns of the explained code smells and the tool could detect all of them. This is, however, a biased accuracy analysis and we have left this as a future work to conduct a large-scale empirical analysis on JavaScript applications, as well as evaluation of the detection accuracy of the tool.

The implemented detection methods might not be entirely accurate, leading to false positives or false negatives. The tool's effectiveness could be influenced by factors such as code complexity, obfuscation, or the specific version of JavaScript being used. Additionally, any improvements or refinements to JSNose may affect the consistency of results, and different configurations of the tool may produce varying outcomes.

\head{Threats to Validity} Our contributions in this research are susceptible to a number of threats to validity, as follows:

\textit{Internal Validity}: One threat to the internal validity of our study is related to the subjective nature of defining and categorizing security code smells. Although we based our definitions on established sources like CWE and OWASP, there is still room for interpretation. We acknowledge that our work may not include all possible \javascript security smells as determining a code as smelly is subjective and based on opinions and experiences \cite{fard2013jsnose}. To mitigate this, we only consider security smells that are mapped to the CWE list, which is a mature community-established list of security weaknesses observed in the real world and have been widely used by the security community. In some cases, however, the mentioned smells could be an actual vulnerability, e.g., using weak cryptography and hard-coded passwords. In addition, we performed the mappings of security code smells to CWEs OWASP top 10 manually, which may introduce error or bias. To mitigate this, we conducted a review process in which each author reviewed the mapping done by others. 

Another potential threat is our proposed metrics and criteria to detect security code smells. HHowever, as suggested in previous work, we believe these metrics and thresholds can effectively identify the described code smells. Our extension to JSNose for detecting the proposed security smells may introduce false positives or false negatives in detection. Some smells may be inaccurately identified as security issues when they pose minimal risk, or valid security smells may not be detected due to limitations in the detection rules. However, the accuracy of the implemented smell detection tool was not the target of this paper. 

\textit{External Validity}: An external validity concerns the generalizability of the proposed security smells beyond vanilla/plain JavaScript such as code written using frameworks or in TypeScript. However, we believe that the studied set of smells are still representative in many JavaScript projects at their core and can be adapted to frameworks and libraries. Moreover, we did not evaluate the accuracy and effectiveness of our implementation on multiple JavaScript applications in different domains at front-end and server-side. We acknowledge that more experimental subjects should be studied and evaluated to support a claim on the accuracy of our detection implementation.

\textit{Construct Validity}: A threat to construct validity in our study is the potential mismatch between the identified code smells and actual security vulnerabilities. While we mapped our smells to well-known security frameworks like CWE and OWASP, the practical impact of these smells on security might vary. To address this, we plan to conduct empirical studies to assess the real-world implications of the detected smells on application security. The definitions and detection methods used for security code smells might not fully capture the underlying vulnerabilities they represent. There could be disagreements or nuances in how researchers or practitioners interpret these concepts.

\textit{Conclusion Validity}: While mapping to CWE and OWASP guidelines provides a theoretical foundation for the security impact of these smells, further empirical studies would strengthen the conclusion that the presence of these code smells can be correlated with higher security risks.

\section{Conclusions and Future Work}
\label{conclusionSection}
We presented 24 \javascript security code smells, mapped them to CWE and OWSP Top 10, explained possible refactoring, and detection mechanism. We extended an existing open source \javascript code smell detector that performs static and dynamic analysis. As noted in \cite{fard2013jsnose}, determining a code as smelly is subjective and based on opinions and experiences. Therefore we support our claims for proposed security smells through references and examples. We acknowledge that our proposed list of \javascript security code smells is not exhaustive. The definitions and detection methods used for security code smells might not fully capture the underlying vulnerabilities they represent. There could be disagreements or nuances in how different researchers or practitioners interpret these concepts.

For future work, we intend to perform an empirical evaluation on \javascript security code smells on real-world applications to determine their prevalence and analyze the relationship between some code metrics and security code smells. Moreover, we plan to produce a dataset of \javascript code snippets with security smells and use it to 1) analyze code smells within similar code, similar to \cite{piran2021vulnerability}, and 2) develop a machine learning-based smell detector. Another possible future direction, similar to what mentioned in \cite{milanifard_2017} for extension of JSNose, can be applying similar methods to proposed security code smells and detection techniques for other implementations of ECMAScript or supersets of JavaScript, such as CoffeeScript and TypeScript, prototype-based programming languages, and languages that support first-class functions.

\bibliographystyle{abbrv}  
\bibliography{paper}

\begin{thebibliography}{10}

\bibitem{Checkmarx}
{Checkmarx: Application Security Tools}.
\newblock \url{https://checkmarx.com/}, 2024.

\bibitem{MITRE}
{Common Weakness Enumeration}.
\newblock \url{http://cwe.mitre.org/data/index.html}, 2024.

\bibitem{CISQ}
{Consortium for Information \& Software Quality (CISQ) Automated Quality
  Characteristic Measures}.
\newblock
  \url{https://www.it-cisq.org/standards/automated-quality-characteristic-measures-maintainability/},
  2024.

\bibitem{ESLint}
Eslint: Find and fix problems in your javascript code.
\newblock \url{https://eslint.org/}, 2024.

\bibitem{JSHint}
Jshint: detect errors and potential problems in your {JavaScript} code.
\newblock \url{https://jshint.com/}, 2024.

\bibitem{JSLint}
Jslint: The {JavaScript} code quality tool.
\newblock \url{http://www.jslint.com/}, 2024.

\bibitem{OWASP}
{OWASP Top 10 Vulnerabilities}.
\newblock \url{https://owasp.org/Top10/}, 2024.

\bibitem{postMessage}
{postMessage API}.
\newblock
  \url{https://developer.mozilla.org/en-US/docs/Web/API/Window/postMessage},
  2024.

\bibitem{Snyk}
{Snyk: Static application security testing}.
\newblock \url{https://snyk.io/product/snyk-code/}, 2024.

\bibitem{SonarSource}
{SonarSource Static Code Analysis for JavaScript}.
\newblock \url{https://rules.sonarsource.com/javascript/}, 2024.

\bibitem{Veracode}
{Veracode: Application Security Tools}.
\newblock \url{https://www.veracode.com/}, 2024.

\bibitem{alenezi2020relationship}
M.~Alenezi and M.~Zarour.
\newblock On the relationship between software complexity and security.
\newblock {\em arXiv preprint arXiv:2002.07135}, 2020.

\bibitem{elkhail2019relating}
A.~A. Elkhail and T.~Cerny.
\newblock On relating code smells to security vulnerabilities.
\newblock In {\em IEEE International Conference on Big Data Security on Cloud
  (BigDataSecurity), IEEE Intl Conference on High Performance and Smart
  Computing,(HPSC) and IEEE intl conference on intelligent data and security
  (IDS)}, pages 7--12. IEEE, 2019.

\bibitem{fowler1999refactoring}
M.~Fowler and K.~Beck.
\newblock {\em Refactoring: improving the design of existing code}.
\newblock Addison-Wesley Professional, 1999.

\bibitem{gadient2021security}
P.~Gadient, M.-A. Tarnutzer, O.~Nierstrasz, and M.~Ghafari.
\newblock Security smells pervade mobile app servers.
\newblock In {\em ACM/IEEE International Symposium on Empirical Software
  Engineering and Measurement (ESEM)}, pages 1--10, 2021.

\bibitem{Ghafari}
M.~Ghafari, P.~Gadient, and O.~Nierstrasz.
\newblock Security smells in android.
\newblock In {\em IEEE International Working Conference on Source Code Analysis
  and Manipulation (SCAM)}, pages 121--130, 2017.

\bibitem{ghaith2012improving}
S.~Ghaith and M.~{\'O}~Cinn{\'e}ide.
\newblock Improving software security using search-based refactoring.
\newblock In {\em International Symposium on Search Based Software
  Engineering}, pages 121--135. Springer, 2012.

\bibitem{gong2015dlint}
L.~Gong, M.~Pradel, M.~Sridharan, and K.~Sen.
\newblock Dlint: Dynamically checking bad coding practices in javascript.
\newblock In {\em International Symposium on Software Testing and Analysis
  (ISSTA)}, 2015.

\bibitem{haas2020static}
R.~Haas, R.~Niedermayr, T.~Roehm, and S.~Apel.
\newblock Is static analysis able to identify unnecessary source code?
\newblock {\em ACM Transactions on Software Engineering and Methodology
  (TOSEM)}, 29(1):1--23, 2020.

\bibitem{Iannone}
E.~Iannone, Z.~Codabux, V.~Lenarduzzi, and A.~De~Lucia.
\newblock Rubbing salt in the wound? a large-scale investigation into the
  effects of refactoring on security.
\newblock {\em Empirical Software Engineering}, 28(4):89, 2023.

\bibitem{lanza2006objectoriented}
M.~Lanza and R.~Marinescu.
\newblock {\em Object-oriented Metrics in Practice: Using Software Metrics to
  Characterize, Evaluate, and Improve the Design of Object-Oriented Systems}.
\newblock Springer, 2006.

\bibitem{meawad2012eval}
F.~Meawad, G.~Richards, F.~Morandat, and J.~Vitek.
\newblock Eval begone! semi-automated removal of eval from javascript programs.
\newblock In {\em ACM International Conference on Object Oriented Programming
  systems languages and applications}, pages 607--620, 2012.

\bibitem{milanifard_2017}
A.~Milani~Fard.
\newblock {\em Directed test generation and analysis for web applications}.
\newblock PhD thesis, University of British Columbia, 2017.

\bibitem{fard2013jsnose}
A.~Milani~Fard and A.~Mesbah.
\newblock {JSNose}: Detecting {JavaScript} code smells.
\newblock In {\em IEEE International Working Conference on Source Code Analysis
  and Manipulation (SCAM)}, pages 116--125. IEEE Computer Society, 2013.

\bibitem{fard2017javascript}
A.~Milani~Fard and A.~Mesbah.
\newblock {JavaScript: The (Un)covered Parts}.
\newblock In {\em IEEE international conference on software testing,
  verification and validation (ICST)}, pages 230--240. IEEE, 2017.

\bibitem{amin:ase15}
A.~Milani~Fard, A.~Mesbah, and E.~Wohlstadter.
\newblock {Generating Fixtures for JavaScript Unit Testing}.
\newblock In {\em IEEE/ACM International Conference on Automated Software
  Engineering (ASE)}, pages 190--200. ACM, 2015.

\bibitem{moha2010decor}
N.~Moha, Y.-G. Gu{\'e}h{\'e}neuc, L.~Duchien, and A.-F. Le~Meur.
\newblock {DECOR}: A method for the specification and detection of code and
  design smells.
\newblock {\em IEEE Transactions on Software Engineering}, 36(1):20--36, 2010.

\bibitem{mumtaz2018empirical}
H.~Mumtaz, M.~Alshayeb, S.~Mahmood, and M.~Niazi.
\newblock An empirical study to improve software security through the
  application of code refactoring.
\newblock {\em Information and Software Technology}, 96:112--125, 2018.

\bibitem{munro2005product}
M.~J. Munro.
\newblock Product metrics for automatic identification of ``bad smell'' design
  problems in {J}ava source-code.
\newblock In {\em Proc. International Symposium Software Metrics}, pages
  15--15. IEEE, 2005.

\bibitem{paramitha2021mining}
R.~Paramitha and Y.~D.~W. Asnar.
\newblock Mining software repository for security smell code review.
\newblock In {\em International Conference on Data and Software Engineering
  (ICoDSE)}, pages 1--6. IEEE, 2021.

\bibitem{piran2021vulnerability}
A.~Piran, C.-P. Chang, and A.~Milani~Fard.
\newblock Vulnerability analysis of similar code.
\newblock In {\em 2021 IEEE 21st International Conference on Software Quality,
  Reliability and Security (QRS)}, pages 664--671. IEEE, 2021.

\bibitem{ponce2022smells}
F.~Ponce, J.~Soldani, H.~Astudillo, and A.~Brogi.
\newblock Smells and refactorings for microservices security: A multivocal
  literature review.
\newblock {\em Journal of Systems and Software}, 192:111393, 2022.

\bibitem{pradel2015typedevil}
M.~Pradel, P.~Schuh, and K.~Sen.
\newblock Typedevil: Dynamic type inconsistency analysis for javascript.
\newblock In {\em IEEE/ACM International Conference on Software Engineering
  (ICSE)}, volume~1, pages 314--324. IEEE, 2015.

\bibitem{rahman2020code}
A.~Rahman, E.~Farhana, and L.~Williams.
\newblock The ‘as code’activities: development anti-patterns for
  infrastructure as code.
\newblock {\em Empirical Software Engineering}, 25:3430--3467, 2020.

\bibitem{rahman2019seven}
A.~Rahman, C.~Parnin, and L.~Williams.
\newblock The seven sins: Security smells in infrastructure as code scripts.
\newblock In {\em IEEE/ACM International Conference on Software Engineering
  (ICSE)}, pages 164--175. IEEE, 2019.

\bibitem{rahman2021security}
A.~Rahman, M.~R. Rahman, C.~Parnin, and L.~Williams.
\newblock Security smells in ansible and chef scripts: A replication study.
\newblock {\em ACM Transactions on Software Engineering and Methodology
  (TOSEM)}, 30(1):1--31, 2021.

\bibitem{rahman2019share}
M.~R. Rahman, A.~Rahman, and L.~Williams.
\newblock Share, but be aware: Security smells in python gists.
\newblock In {\em IEEE International Conference on Software Maintenance and
  Evolution (ICSME)}, pages 536--540. IEEE, 2019.

\bibitem{saboury2017empirical}
A.~Saboury, P.~Musavi, F.~Khomh, and G.~Antoniol.
\newblock An empirical study of code smells in javascript projects.
\newblock In {\em 2017 IEEE 24th international conference on software analysis,
  evolution and reengineering (SANER)}, pages 294--305. IEEE, 2017.

\bibitem{sariman2016novel}
G.~Sar{\i}man and E.~U. Kucuksille.
\newblock A novel approach to determine software security level using bayes
  classifier via static code metrics.
\newblock {\em Elektronika ir Elektrotechnika}, 22(2):73--80, 2016.

\bibitem{mccabe}
B.~Schneier.
\newblock More complex = less secure, miss a test path and you could get
  hacked.
\newblock Technical report, McCabe Software, 2012.
\newblock
  \url{http://www.mccabe.com/pdf/MoreComplexEqualsLessSecure-McCabe.pdf}.

\bibitem{shin2010evaluating}
Y.~Shin, A.~Meneely, L.~Williams, and J.~A. Osborne.
\newblock Evaluating complexity, code churn, and developer activity metrics as
  indicators of software vulnerabilities.
\newblock {\em IEEE transactions on software engineering}, 37(6):772--787,
  2010.

\bibitem{shin2008empirical}
Y.~Shin and L.~Williams.
\newblock An empirical model to predict security vulnerabilities using code
  complexity metrics.
\newblock In {\em Proceedings of the Second ACM-IEEE international symposium on
  Empirical software engineering and measurement}, pages 315--317, 2008.

\bibitem{simon2001metrics}
F.~Simon, F.~Steinbruckner, and C.~Lewerentz.
\newblock Metrics based refactoring.
\newblock In {\em Proc European Conference on Software Maintenance and
  Reengineering (CSMR)}, pages 30--38. IEEE, 2001.

\bibitem{singh2020long}
R.~Singh, A.~Bindal, and A.~Kumar.
\newblock Long method and long parameter list code smells detection using
  functional and semantic characteristics.
\newblock {\em International Journal of Recent Technology and Engineering
  (IJRTE)}, 8(6):245--253, 2020.

\bibitem{sultana2020examining}
K.~Z. Sultana, Z.~Codabux, and B.~Williams.
\newblock Examining the relationship of code and architectural smells with
  software vulnerabilities.
\newblock In {\em 2020 27th Asia-Pacific Software Engineering Conference
  (APSEC)}, pages 31--40. IEEE, 2020.

\bibitem{zhang2020detecting}
M.~Zhang and W.~Meng.
\newblock Detecting and understanding javascript global identifier conflicts on
  the web.
\newblock In {\em Proceedings of the 28th ACM Joint Meeting on European
  Software Engineering Conference and Symposium on the Foundations of Software
  Engineering (FSE)}, pages 38--49, 2020.

\end{thebibliography}

\end{document}